\documentstyle[aps,prl,epsf]{revtex}

\newcommand{\be}{\begin{equation}}
\newcommand{\ee}{\end{equation}}
\newcommand{\bbet}{\mbox{\boldmath $ \beta $}}
\newcommand{\bdelbet}{\mbox{\boldmath $ \delta \beta $}}

\newcommand {\fig}[1]{fig.$\;$\ref{#1}}
\newcommand {\Fig}[1]{Fig.$\;$\ref{#1}}

\begin{document}
\advance\textheight by 0.5in
\advance\topmargin by -0.2in
\draft
\twocolumn[\hsize\textwidth\columnwidth\hsize\csname@twocolumnfalse%
\endcsname

\title{Structure and Strength of Dislocation Junctions: An Atomic Level Analysis}

\author{D. Rodney and R. Phillips}

\address{Division of Engineering, Brown University, Providence, RI 02912}

\date{\today}

\maketitle

\begin{abstract}
The quasicontinuum method is used to simulate three-dimensional 
Lomer-Cottrell junctions both in the absence and in the presence
of an applied stress. The simulations show that this type of
junction is destroyed by an unzipping 
mechanism in which the dislocations that form the junction
are gradually pulled apart along
the junction segment. The calculated critical stress needed for
breaking the junction is comparable to that predicted
by line tension models. The simulations also demonstrate
a strong influence of the initial dislocation line directions
 on the breaking mechanism, an effect that is
 neglected in the macroscopic treatment of the hardening
 effect of junctions.
\end{abstract}
]

The mechanical properties of a wide range of materials
can be traced in part to the motion and interaction of dislocations.
Dislocation junctions serve as one class of obstacles to dislocation
motion. From the standpoint of work hardening, junctions have been
implicated as a primary contributor to the observed
increase in the flow stress with increasing dislocation
density \cite{basinski}.
They are one example of   "dislocation
chemistry" in which dislocations can join and dissociate to form
new segments. Such reactions involve both atomic and elastic
effects since in the 
core regions there are substantial atomic rearrangements, 
while at larger distances the dislocations still interact elastically.

Until now, the vast majority of information
concerning dislocation junctions has been
obtained either from macroscopic hardening 
experiments \cite{franciosi} or 
models deriving from elasticity theory \cite{saada,schoeck}.
On the other hand, evidence has been mounting
for decades that in certain circumstances dislocation
core effects play a critical role in determining
the actual behavior of  materials\cite{vitek},
and recent calculations have shown that atomic level
insights can be gained with respect to junctions as
well \cite{zhou}.
 The calculations
undertaken here attempt to account for both the
long range elastic interactions and the detailed atomic
level geometries that give junctions their overall behavior.
Though we have made a systematic study of a number of
dislocation junctions found in fcc materials, we concentrate here 
only on the case of the Lomer-Cottrell junction.
Our reason for limiting the discussion  is the  presumed importance of
the Lomer-Cottrell junction in governing the
hardening of fcc materials as deduced from
experimental analyses \cite{franciosi}.

Our ambition in the pages that follow is to examine the atomic-level
structure of the Lomer-Cottrell junction both in
the absence and presence of an externally applied stress.
The response of a junction to an applied stress is of prime
importance since, in macroscopic descriptions of hardening \cite{kocks},
obstacles to dislocation motion such as junctions are 
represented by parameters derived from the stress needed to
force the dislocations across the obstacles of interest. 

Although such an investigation demands an atomic level
calculation, it is also evident that three-dimensional
calculations of this sort  require an enormous computational investment.  As an
alternative to full-scale atomistic simulations,
we have used the quasicontinuum method \cite{shenoy} as the primary
vehicle for our analysis.  By linking finite element
discretization with atomic level calculations it is
possible to carry out simulations in which only small
neighborhoods of the dislocation slip planes
 are treated with
full atomic resolution, while in the far fields a coarser finite
element mesh is exploited.  In order to make sure
that there were no mesh effects, we performed 
simulations on the same geometries with varying atomistic regions.

\begin{figure}
\narrowtext \centerline{\epsfxsize=1.5truein 
\epsfbox{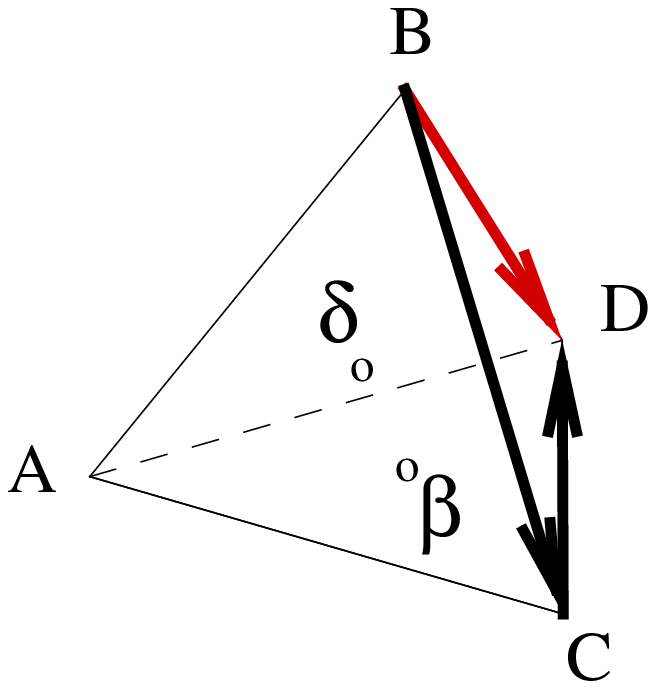} \epsfxsize=1.5truein
\epsfbox{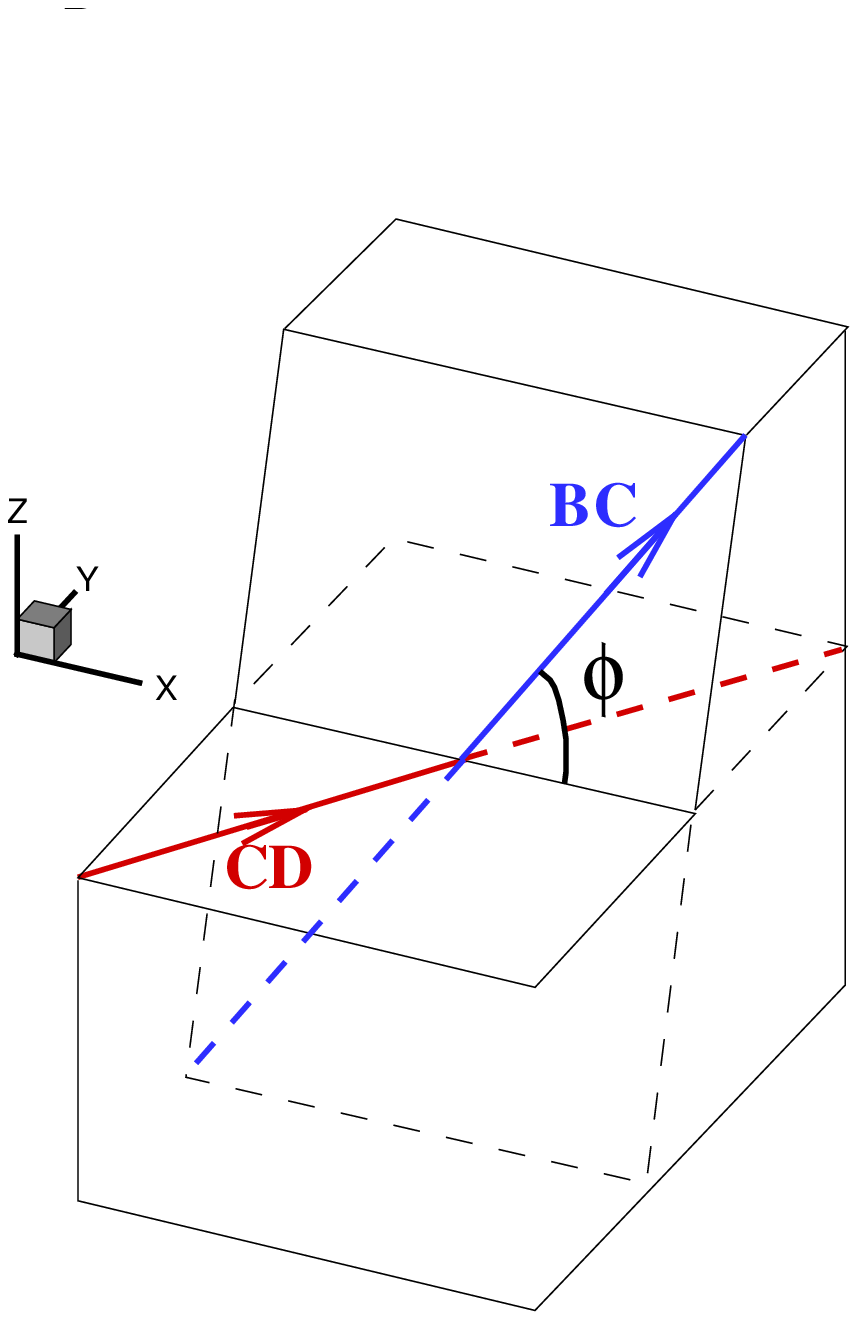}}
\centerline{(a)\hspace{1.truein}(b)}
\medskip
\caption{(a) Thompson tetrahedron illustrating the Burgers
vectors of the dislocations of interest.  (b) Schematic section
of the initial simulation cell along the glide planes
of the dislocations.
The two dislocations have the same length, $2l$, and make the same
angle $\phi$ with the line of intersection between the glide planes.}
\label{schematic}
\end{figure}

The basic elements of our simulations are carried out as follows.
The geometry is best understood
by appealing to \fig{schematic}.  The
Thompson tetrahedron shown in \fig{schematic}(a) is used to characterize
the possible dislocations that can form in an fcc
material. We consider the case of two interacting 
$ a / 2  \langle 110 \rangle \{111 \}$
dislocations shown schematically in \fig{schematic}(b). The
first dislocation lives in the horizontal ACD plane also labeled $\beta$ in
\fig{schematic}(a) and has a Burgers vector equal to the vector $\bf{CD}$.
 We imagine that this dislocation
interacts with a partner dislocation in the inclined slip plane ABC
(labelled $\delta$) with
a Burgers vector $\bf{BC}$. The Burgers vector of the junction
segment is  the sum of the initial vectors. From the Thompson tetrahedron,
it is obvious that this vector is $\bf{BD}$ which  is in
neither of the initial glide planes. The significance of this observation
is that the resulting junction is expected to be sessile since it 
can glide in neither of the 
relevant slip planes.  As a result, this type of 
junction is expected to impede the motion of the  
dislocations and therefore is a source of hardening. 

To carry out the simulations, the dislocations are
installed in a simulation cell by virtue of their linear elastic
fields.  All the nodes of the mesh are
displaced in accordance with the superposition of the Volterra
fields due to the two participating dislocations.  Typical cell
dimensions are of the order of $(300\AA)^3$. 
 A typical simulation demands around
$1.5 \times 10^5$ representative atoms, in contrast with the $1.5 \times 10^6$ atoms
that would be required to carry out a conventional
lattice statics simulation with  the same geometry.
The dislocations are initially straight and undissociated. 
We have chosen a highly symmetric
geometry in which the angle between the dislocations and the
line of intersection of the two glide planes is $\phi = 60^o$ and the
two dislocations have the same length $2l$. This idealized geometry
is similar to that envisaged in elastic models for junction
formation \cite{saada,schoeck} and  therefore lends itself to direct comparison
with such models.
Simulations were performed with dislocation lengths
 equal to $300$ and $400 \AA$.
 
The initial separation between the dislocations is
$6\AA$ leading to a strong mutual interaction.
The atoms on the boundary of the computational
cell are held fixed in accordance with their initial elastic
displacements. Because of these rigid boundary conditions,
the dislocations are pinned at their ends. The equilibrium
structure is then determined as the energy minimizing configuration
(obtained by conjugate gradients) associated with
the boundary conditions of interest. For the purposes of the
present analysis, we have exploited the embedded atom (EAM) 
potentials of Ercolessi and Adams \cite{ercolessi}
developed to simulate aluminum crystals.

In order to garner some idea of the role of finite size 
effects, it is useful to estimate the magnitude of the
dislocation densities corresponding to
the geometry present  in our simulations.
We can imagine that the dislocations are pinned on the surfaces
of the simulation cell because they take part in other junctions
at these points. The distance between the junctions in
this imaginary network is therefore
close to $150\AA$. 
The classical relation coupling the mean separation between
pinning points and the density of dislocations yields
$\rho \sim 1/l^2 \sim 10^{15} m^{-2}$  which corresponds to the upper
limit for densities in heavily worked materials. 

For the purposes of  examining the action of an applied stress and to 
force the horizontal dislocation in plane $\beta$ 
to glide through the inclined one, we have carried out 
a sequence of quasistatic calculations in which the simulation cell itself
is subjected to incremental shears.
A homogeneous shear $\gamma_{yz} = \partial u_y /\partial z
= constant$ is applied 
 in increments  of $0.27\%$.
In the
linear elastic approximation, this deformation corresponds 
to resolved shear stresses equal to 
 $\frac{\sqrt 3}{2} \mu \gamma_{yz}$ on the horizontal dislocation
 and $\frac{7}{6 \sqrt 3} \mu \gamma_{yz}$ on the inclined one.
 As will be shown below, the strains at which the junctions
 break are of order of $2\%$, which may well exceed the limits
 of validity of the linear elastic approximation suggested
 above.
   For each load step, the energy is
 minimized, with the result that a series of
 equilibrium states under increasing applied stress are obtained.
The results
of the simulations are visualized by plotting
only those atoms whose energy exceeds that of atoms 
in the middle of a stacking fault.

\begin{figure*}
\narrowtext \centerline{
\epsfxsize=1.5truein \epsfbox{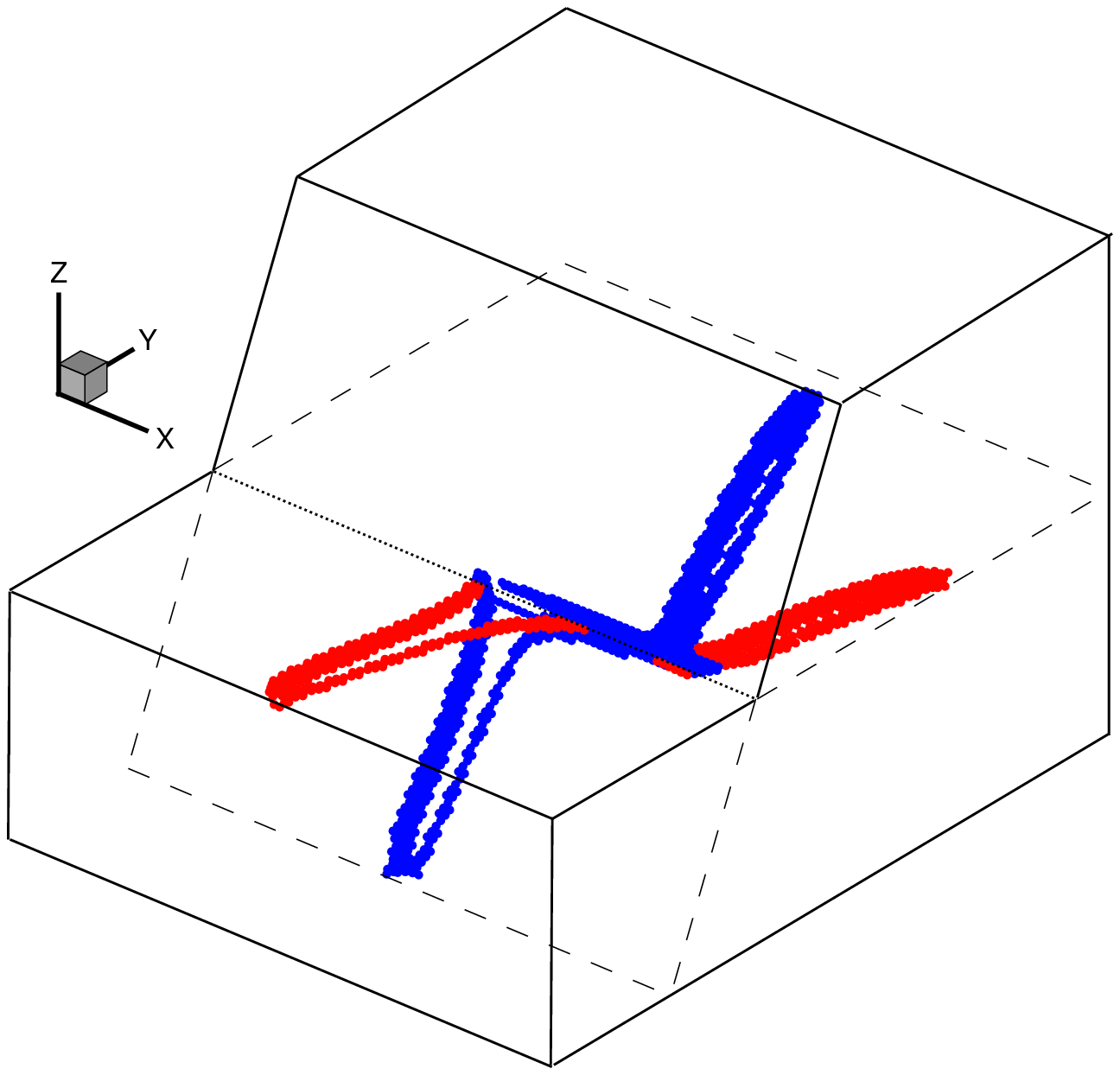}\hspace{.2truein}
\epsfxsize=1.5truein \epsfbox{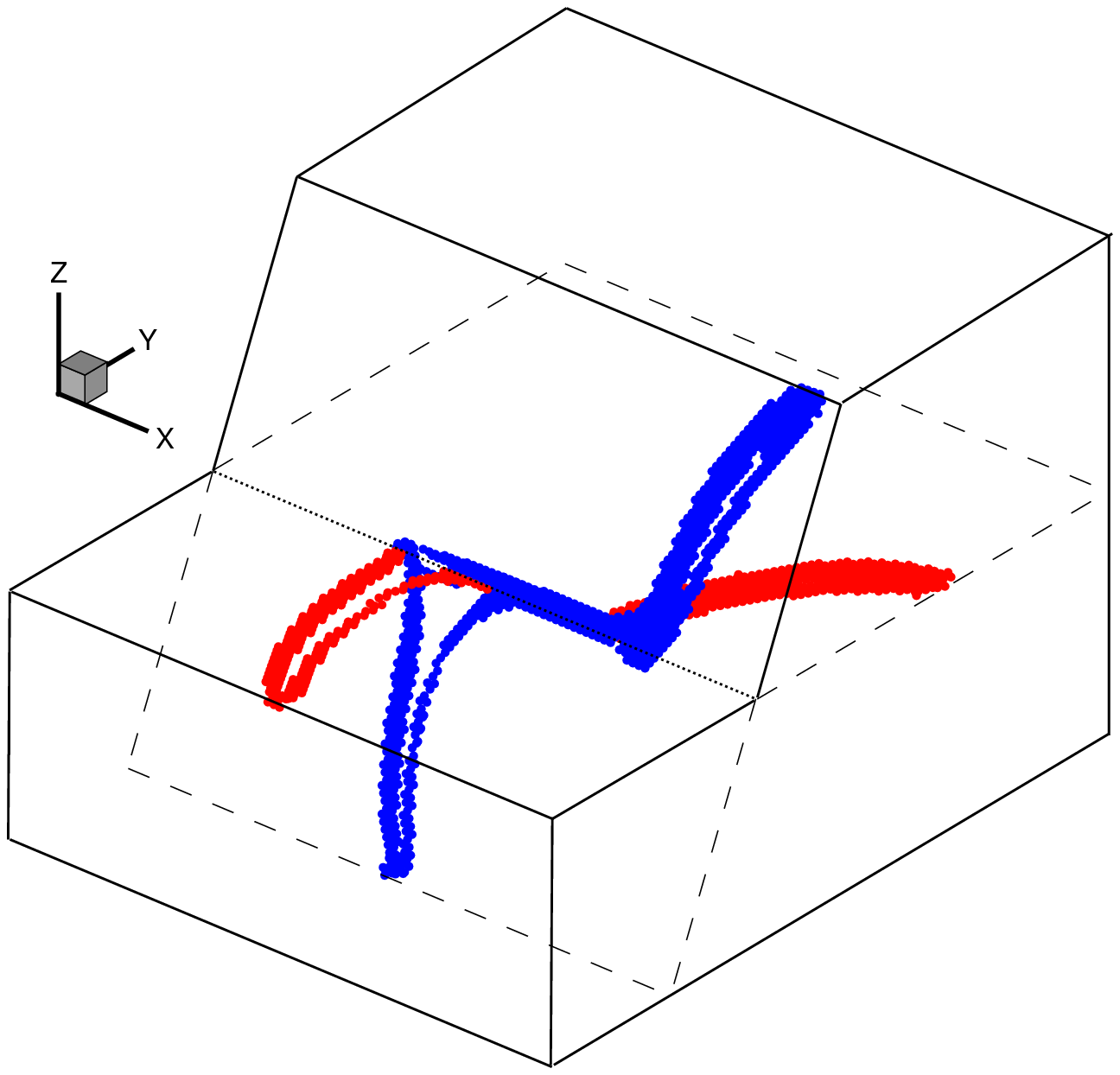} }
\centerline{(a)\hspace{2.truein}(b)}
\centerline{
\epsfxsize=1.5truein \epsfbox{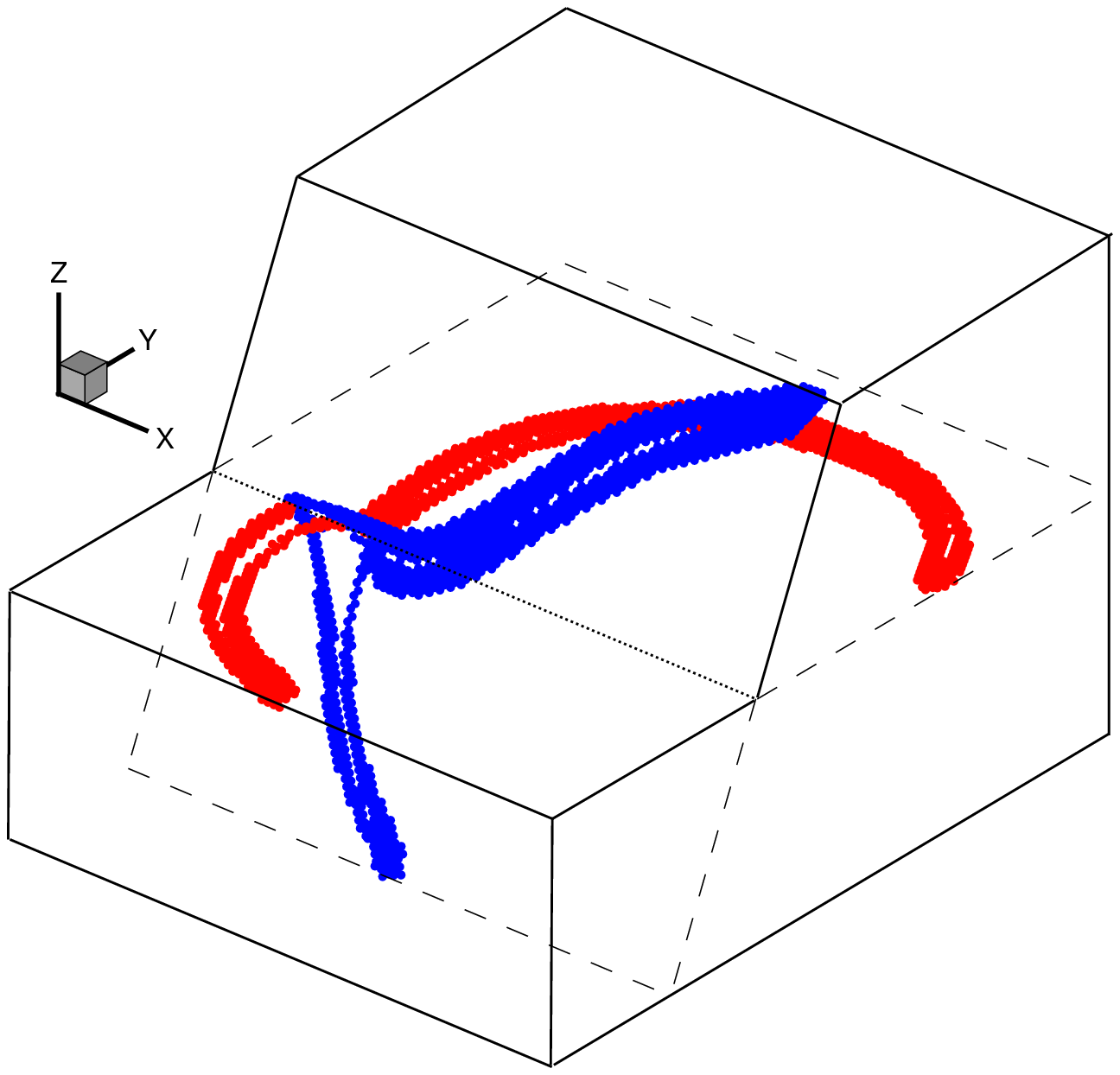}\hspace{.2truein}
\epsfxsize=1.5truein \epsfbox{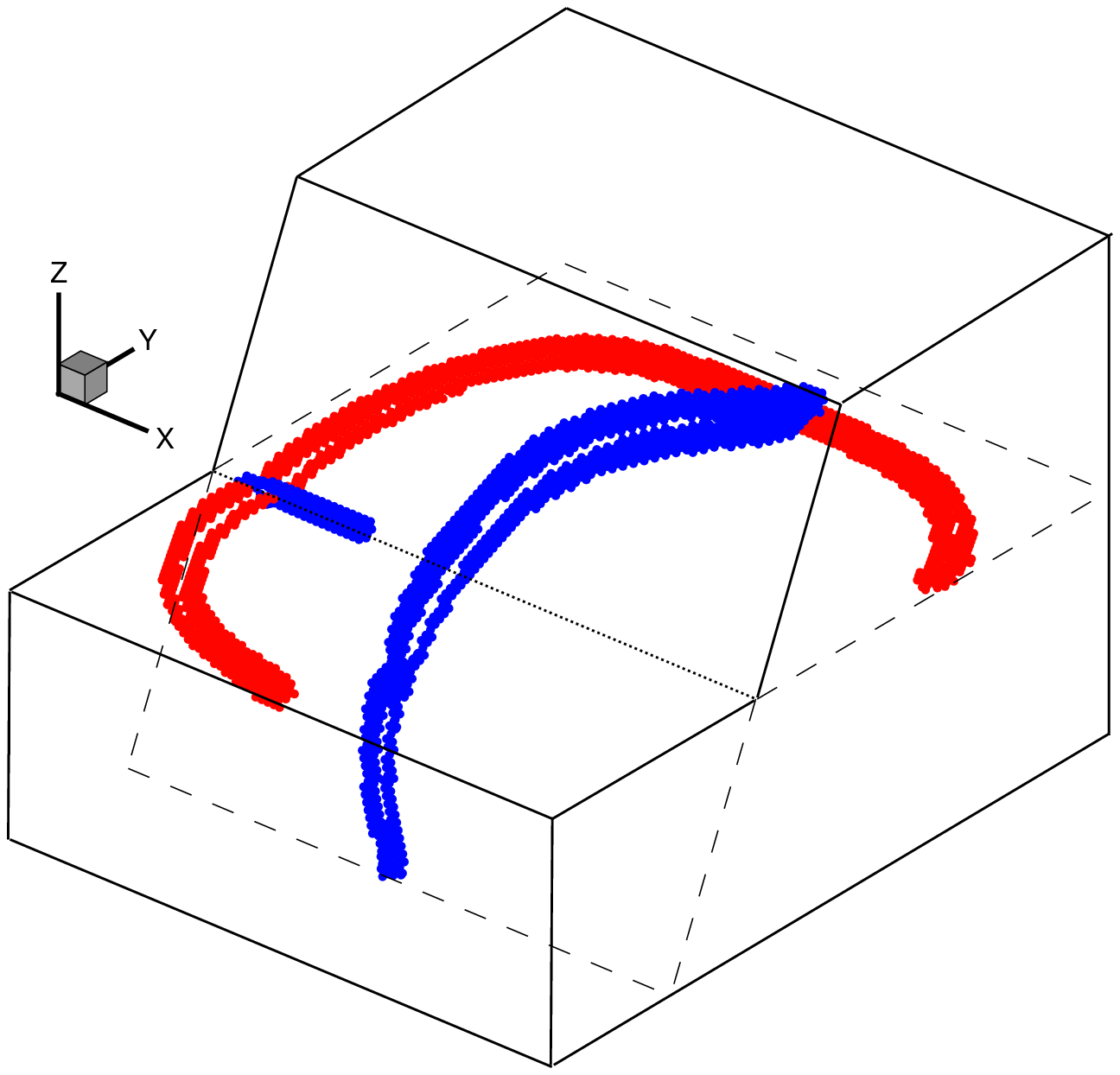} }
\centerline{(c)\hspace{2.truein}(d)}
\medskip
\caption{Sequence of snapshots of the junction geometry under
increasing stress. (a) Zero applied stress, (b) stress is $0.011\mu$, (c) stress is $0.018\mu$ just before the junction breaks, (d) stress is $0.018\mu$ at the end of the simulation.}
\label{sixty}
\end{figure*}

\Fig{sixty} shows the disposition of the junction at four stages
of deformation.
\Fig{sixty}(a) illustrates the structure in the absence of any applied stress,
with the result that a junction is formed along the line 
of intersection between the glide planes of the initial dislocations.
The four arms of the
junction are dissociated in their glide planes 
as is evidenced by the presence of
two segments on each arm which are Shockley partials.
 The Burgers vectors of all of
the dislocation segments
can be computed from the Burgers vectors of the initial
segments and the vectors
of the stacking faults. 
The junction segment itself  is built up of two parts.
A stair-rod Lomer-Cottrell segment with a 
Burgers vector \bdelbet~
 of the type $ a / 6 \langle 110 \rangle $ (see \fig{schematic}(a))
forms one edge of the  extended node on the left side of the junction.
The remaining part of the junction segment is a Lomer dislocation with Burgers vector
$\bf{BD}$ as expected from the crystallographic
analysis given above. The nature of this dislocation is also evidenced
by its five-fold symmetrical core which is seen on closer
inspection of the atomic region.
Unlike the left hand node, the right hand node is
pointlike and is the meeting point of the constricted segments. The size of
the extended node is independent of the initial dislocation length
and of the initial angle $\phi$ and is therefore dictated by the
equilibrium of the incoming partials only. On the other hand, the
simulations show that
the length of the Lomer segment increases with decreasing initial
angle $\phi$ and increasing dislocation length. This latter result
was expected on the basis of elastic models \cite{saada}.

The detailed structure of the junction itself involves other
interesting features as well. 
Kinematic arguments \cite{heidenreich} predict that dislocations acquire a
step after passing through each other, whose height is equal to
the Burgers vector of the dislocation being crossed.  
In the present setting, even in the initial stress free configuration,
this discontinuity is present in the sense that arms of
the junction which once formed a single dislocation, for example
the two horizontal arms, already live on {\it different} 
parallel $ \{111\} $ planes.
 The origin of this structure is the discontinuity of the
deformation across both glide planes behind the dislocations.

The results of our simulations can be considered in light
of both earlier experimental and theoretical work. The global
structure as well as  the fact that the junction is along the
line of intersection between glide planes agrees with the
assumed structures of the elastic models.
Moreover the calculated junction is
compatible with TEM observations
\cite{amelynckx,karnthaler} made on AlCu alloys
with low stacking fault energies. The major difference between our results
and those found in the experiments is the existence in the 
simulations of the undissociated  Lomer segment.
In the experiments, the entire
length of the junction segment is dissociated.
On the other hand, this effect can be rationalized by appealing to
the relatively high stacking fault energy associated with the
EAM potentials used here.

The simplest elastic model  (based on a uniform line
tension approximation \cite{saada})
 allows for a three-dimensional generalization of
Frank's rule (which states that in two dimensions
two dislocations will combine if 
${\bf b_1 \cdot b_2} < 0$)
  such that
a junction will form not only if the Burgers vectors satisfy a similar
condition, but also if the initial angle $\phi$ is less than some critical value.
In fact, the geometry considered above corresponds to an initial
angle which is precisely the critical angle separating those 
for which a junction forms
from  those in which there is no junction formation.
An important future direction for this type of analysis is to make
a systematic study of the tendency towards junction formation
as a function of this initial angle.

In addition to shedding light on the propensity for stable
junction formation, elastic models can also be used in
order to examine the stability of such junctions under the
action of an applied stress.
The models predict that $(1)$ 
the arms bow out under the applied stress, $(2)$ the junction
translates along the line of intersection and $(3)$ its length
decreases monotonically with increasing stress 
via an 'unzipping' mechanism.   Unzipping takes
place as the two dislocations forming the junction 
segment are pulled apart under the
action of the applied stress.
The models also predict that,
depending on the dislocation line directions, there exists either
a stress where the length of the junction becomes zero or a stress
above which the junction is unstable. The junction
is destroyed if either of these critical stresses is reached. 
Line tension models  show that the critical stress to break
the junction decreases as the length of
the junction arms increases.
 In particular, with a uniform line tension, in the configuration
 envisaged in the simulations, the critical resolved
shear stress averaged over all possible
line directions is estimated to be 
$ \sim 0.5 \mu b / l$ where l is the length of the dislocation arms.

The mechanism for destruction of the
junction observed in the simulations is similar to that envisaged in
the elastic
models. As can be seen in \fig{sixty}(b), (c) and (d),
each arm bows out  as a result of the non zero resolved shear stresses on
both of the dislocations.
With increasing stress, the junction segment 
translates along the line of intersection between 
glide planes but does not bow out itself,  confirming the fact that 
the junction is a sessile lock.
Between $\sigma = 0$ and $\sigma = 0.011 \mu$,
 the length of the Lomer segment increases in contrast with the
 predictions of the elastic models  which predict that
 the length of the junction segment decreases monotonically
 with increasing stress.  After this point,
with increasing stress, the length of the Lomer
segment decreases until the stress reaches a value of
$\sigma = 0.018 \mu$, at which point 
the Lomer segment  disappears in keeping with
the unzipping mechanism in which the junction length
decreases until it vanishes altogether. 
The net result of the breaking of the junction is that the
dislocation in the horizontal glide plane is now a fully 
bowed out single dislocation, while the
dislocation on the inclined plane snaps back to a more nearly
straight configuration because of the lower resolved shear stress
associated with that dislocation.

Other consequences of the passage of the horizontal dislocation 
through the other can be observed in the simulations. In particular, a
tube of energetic atoms forms behind the horizontal dislocation.
The simulations show  that this is a tube of vacancies which originate from  the step
that forms on the horizontal dislocation when the junction breaks.
As mentioned earlier, the arms of the dislocation live on different
$\{111\}$ planes with the consequence that when they join to reconstruct
the initial dislocation, the latter acquires a jog whose 
height is equal to the interplanar spacing. 
It can be seen from the simulations that the jog on the horizontal dislocation
is sessile. As the dislocation expands,
it leaves behind a dislocation dipole which is
 equivalent to a tube of vacancies since its height is only an interplanar
spacing. 
As no temperature effects 
are accounted for in the simulations, 
the vacancies cannot diffuse away and remain along the tube.
Similarly a jog forms on the inclined dislocation. 
In this case, the jog is constricted and glissile. Its motion along its glide
plane (the BCD plane of \fig{schematic}(a)) 
can be followed in the simulations.

From a quantitative viewpoint, our simulations
reveal that the critical resolved shear stress for breaking the junction
is $\sigma_c = 0.017 \mu = 0.8 \mu b / l$ where $l$ is the 
dislocation arm length. This value  
is of the same order as those determined on the basis of the elastic models.
Note however that there exists important differences between 
the calculated configurations and what was expected from
the models. For example, the deformations observed are much larger
than the limit of validity of the line tension approximation. Moreover
the wavy structure of the inclined arm in \fig{sixty}(c) cannot be
explained in a line tension approximation and is due
to a core effect of the junction.

\begin{figure}
\narrowtext
\centerline{\epsfxsize=2.truein \epsfbox{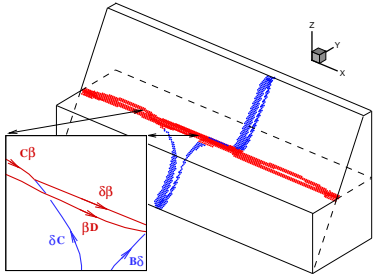}}
\medskip
\caption{Structure of a second Lomer-Cottrell junction in the 
absence of applied stress. The inset is a close-up showing
schematically the configuration of the extended node. The vectors
refer to the Burger vectors of the dislocation segments}
\label{aligned}
\end{figure}

The nature of a junction geometry and the mechanisms
whereby the junction is destroyed are
significantly affected by the line direction of the parent dislocations
as well.  This point is made more clear by appeal to \fig{aligned} in
which we show the results for exactly the same junction type as considered
earlier, but now with a horizontal dislocation initially along the
line of intersection between glide planes.

A particularly interesting feature of the second geometry is the distinct
mechanism that is associated with the destruction of the junction. As
before, a junction forms along the line of intersection between glide 
planes. 
 The right end  is constricted while the  one on the left is extended
with the difference 
 that in this case one of the Shockley partials is
alongside the junction segment itself. Their separation is small
and no perfect stacking fault forms in this region. The inset
in \fig{aligned} is a schematic view of the configuration
near the extended node which shows both the junction segment
$\bdelbet$ and the $\bbet \bf{D}$ Shockley partial. 
In the presence of the applied stress, the arms bow out as previously
described. After the length of the Lomer segment increases, it begins to unzip,
 but  becomes unstable before
it has completely disappeared. The applied stress pushes the Shockley
partial  mentioned above
 against the Lomer-Cottrell segment. The junction
becomes unstable when the stress becomes high enough to force the
"re-association" of the $\bdelbet$ and $\bbet \bf{D}$ segments. 
Note however that this instability 
is not of the same nature as the one predicted by the elastic models. Here,
its origin is the equilibrium of the extended node and
is related to the dissociation of the dislocations which is not taken into
account in the models. Once the junction is destroyed, jogs appear on 
both dislocations. 

The results of our analysis demonstrate that much of the intuition built
up concerning junctions on the basis of line tension models
are borne out qualitatively in the context of
detailed atomic level analyses.  On the other hand, from a quantitative 
standpoint, we have found that the stability of junctions both in
the absence and in the presence of an applied stress 
can be quite different from the simplest elastic estimates. 
The logical extension of the work begun here will be to effect a
series of systematic studies on the role of the initial angle
between the participating dislocations in governing
the propensity for junction formation and  the role of the
junction arm lengths in determining the critical stress for
breaking the junction.  In addition, it remains an outstanding
challenge to continue efforts \cite{bulatov} to link the type of analysis made here
to higher level models of dislocation motion in which 
dislocations are presumed to interact with obstacles to slip that
are characterized by nothing more than a breaking force such as the one
evaluated in the present work.
   
We acknowledge many helpful and illuminating discussions with
J. Bassani, V. Bulatov, G. Canova, M. Fivel,
M. Ortiz, G. Saada and V.B. Shenoy.  In addition,
we gratefully acknowledge the support of Electricit\'e de France,
the NSF supported MRSEC at Brown University, and the Caltech ASCI
program.

\end{document}